\documentclass{aa}

\usepackage{graphicx}
\usepackage{times}
\usepackage{dcolumn}
\usepackage{natbib}

\bibpunct{(}{)}{;}{a}{}{,}
\newcolumntype{.}{D{.}{.}{-1}}

\begin{document}


\title{HE 0047$-$1756: A new gravitationally lensed double QSO
          \thanks{Based in part on observations obtained with the Baade 6.5-m
                  telescope of the Magellan Consortium.
                  Based in part on observations collected at the
                  German-Spanish Astronomical Center, Calar Alto, 
                  operated jointly by Max-Planck Institut f\"ur Astronomie 
                  and Instituto de Astrofísica de Andalucia (CSIC).
                  }}

\author{Lutz Wisotzki\inst{1,2} 
        \and 
        Paul L. Schechter\inst{3}
        \and
        Hsiao-Wen Chen\inst{3}
        \and
        Douglas Richstone\inst{4}
        \and
        Knud Jahnke\inst{1}
        \and
        Sebastian F. S\'{a}nchez\inst{1}
        \and
        Dieter Reimers\inst{5}
        }

\authorrunning{L. Wisotzki et al.}
\titlerunning{HE~0047$-$1756: A new double QSO}

\institute{%
           Astrophysikalisches Institut Potsdam, An der Sternwarte 16, D-14482
           Potsdam, Germany, email: lwisotzki@aip.de
           \and
           Universit\"at Potsdam, Am Neuen Palais 10, 14469 Potsdam, Germany, 
           \and
           Center for Space Research, Massachusetts Institute of
           Technology, Cambridge, MA 02139, USA
           \and
           Dept.\ of Astronomy, University of Michigan, 830 Dennison, 501 East
           University Avenue, Ann Arbor, MI 48109, USA
           \and
           Hamburger Sternwarte, Universit\"at Hamburg, Gojenbergsweg 112, 
           21029 Hamburg, Germany
          }

\date{Draft \today}

\abstract{The quasar HE 0047$-$1756, at $z=1.67$, is found to be split 
into two images $1\farcs44$ apart by an intervening galaxy acting 
as a gravitational lens.  The flux ratio for the two components is 
roughly 3.5:1, depending slightly upon wavelength.  
The lensing galaxy is seen on images obtained at 800~nm and 2.1~$\mu$m;
there is also a nearby faint object which may be responsible for some shear.
The spectra of the two quasar images are nearly identical, but the 
emission line ratio between the two components scale differently
from the continuum. Moreover, the fainter component has a bluer 
continuum slope than the brighter one. We argue that these small
differences are probably due to microlensing.
There are hints of an Einstein ring emanating from the
brighter image toward the fainter one.
}

\maketitle

\section{Introduction}

Every quasar lensed by an intervening galaxy presents opportunities to
study the structure of quasars, lensing galaxies and the intervening
intergalactic medium.  Every year sees the discovery of new
systems\footnote{Nine discoveries were reported in the 2003 postings on the
\texttt{xxx.arXiv.org} preprint server} and
followup studies that take advantage of the peculiar circumstances 
of each new system. 

In this paper we report the discovery a new gravitational lens,
producing two images of the quasar HE~0047$-$1756.  It is the second
system found in the course of an ongoing survey for the lensing of
quasars using the Magellan consortium 6.5-m telescopes on Cerro Las
Campanas \citep{wisotzki*:02:HE0435}.  We present results 
from optical and near-IR images obtained with these telescopes,
plus integral field data obtained at Calar Alto observatory.

\section{Observations}    \label{sec:obs}

\subsection{Identification}    \label{sec:spec}

HE~0047$-$1756 was first discovered
in the course of the Hamburg/ESO bright quasar survey 
(HES; \citealt{wisotzki*:00:HES3}). The survey uses
digitised objective prism spectra to perform a largely automated 
QSO search covering the entire southern extragalactic sky. 
HE~0047$-$1756 was identified in this survey as a 
high probability QSO candidate with redshift $z = 1.66$. However, its apparent
magnitude of $B_J = 16.98$ (measured in the \emph{Digital Sky Survey}
linked to HES photometry) put it below the flux threshold for
systematic follow-up spectroscopy in that particular field, 
and the object remained listed in the archive as a 
high probability QSO candidate
with formal redshift confirmation still pending. 

The Magellan lens survey targets apparently bright quasars at 
substantial redshifts. Because of magnification bias 
\citep{turner:80:GLQ}, such objects have an enhanced 
probilility of being affected by lensing. Although HE~0047$-$1756
was not yet a confirmed QSO, the nature of the object 
was thought to be sufficiently well determined to include it, 
together with several similar cases, in the target list.

\begin{figure*}[tb]
\setlength{\unitlength}{1mm}
\begin{picture}(180,70)
\put(00,36){\includegraphics*[width=34mm]{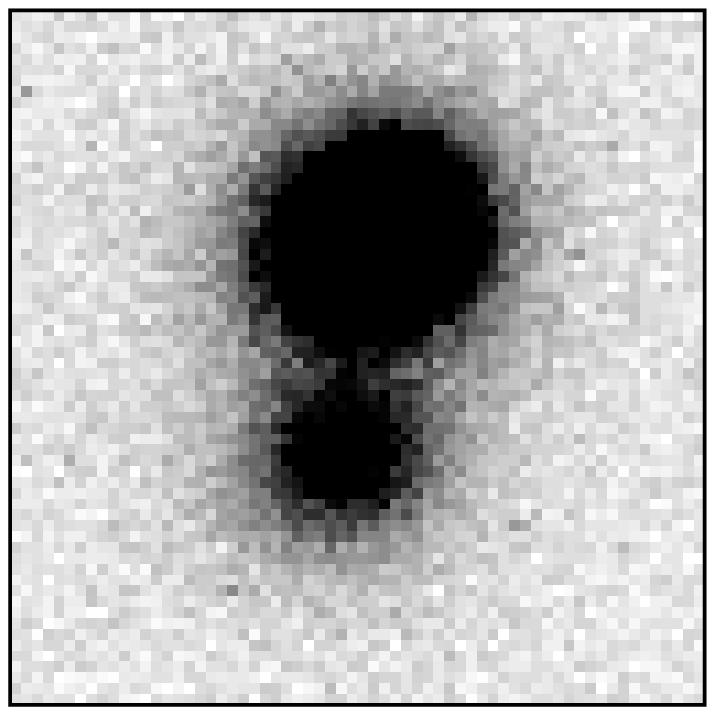}}
\put(03,39){\Large$u'$}
\put(36,36){\includegraphics*[width=34mm]{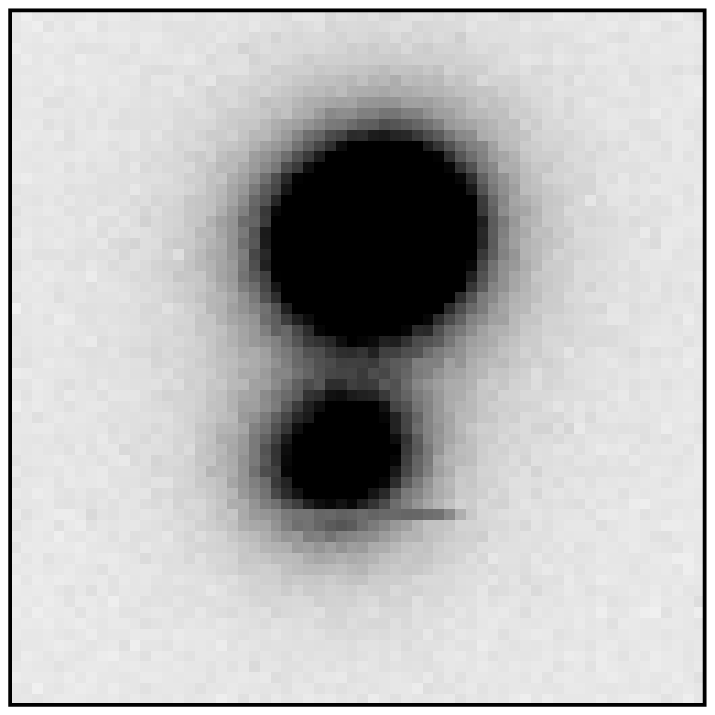}}
\put(39,39){\Large$g'$}
\put(72,36){\includegraphics*[width=34mm]{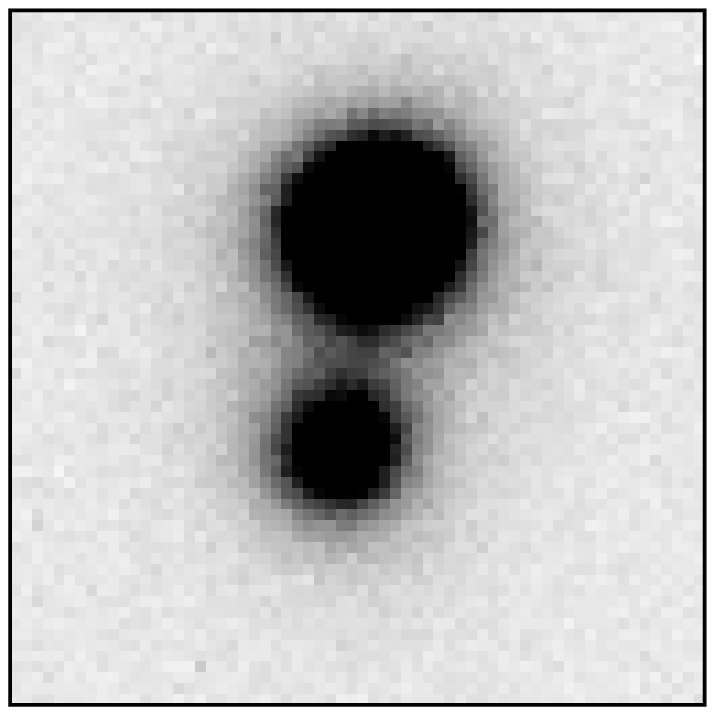}}
\put(75,39){\Large$r'$}
\put(108,36){\includegraphics*[width=34mm]{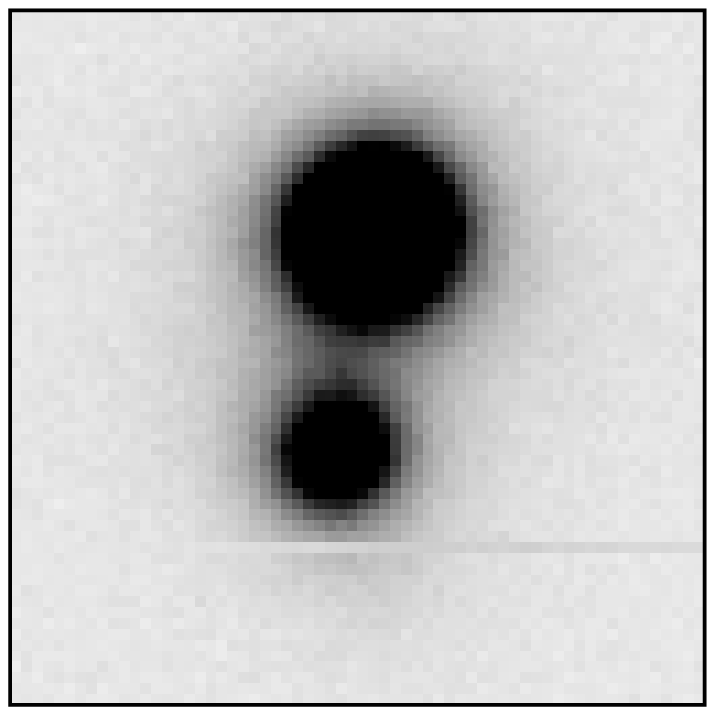}}
\put(111,39){\Large$i'$}
\put(144,36){\includegraphics*[width=34mm]{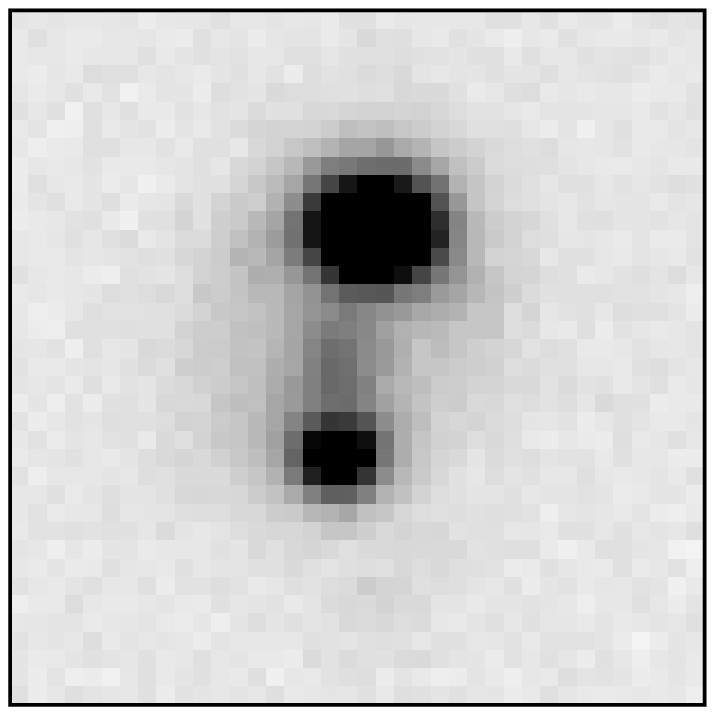}}
\put(147,39){\Large$K_s$}
\put(00,00){\includegraphics*[width=34mm]{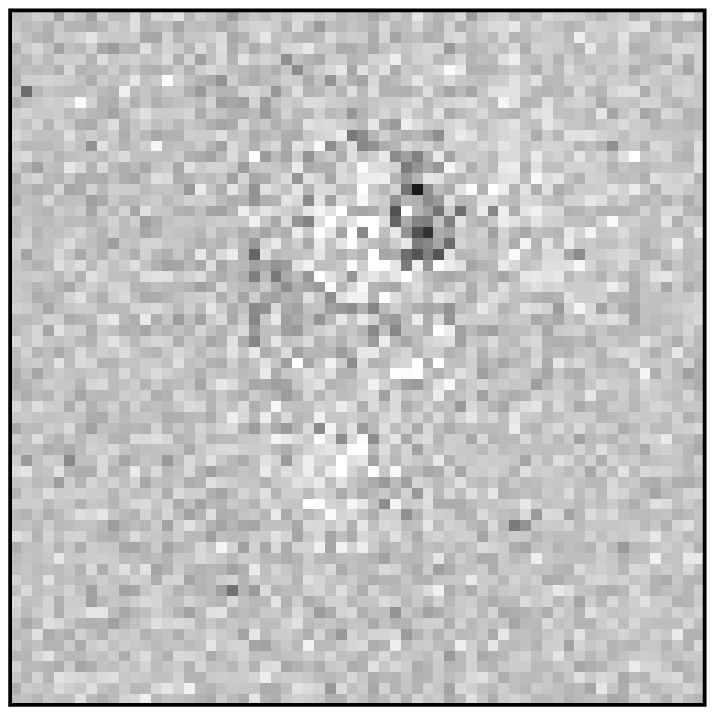}}
\put(03,03){\Large$u'$}
\put(36,00){\includegraphics*[width=34mm]{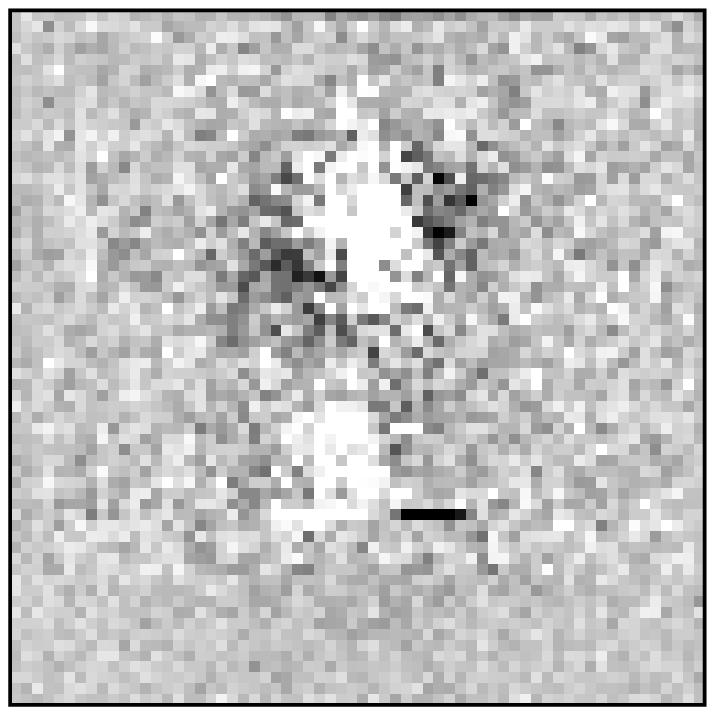}}
\put(39,03){\Large$g'$}
\put(72,00){\includegraphics*[width=34mm]{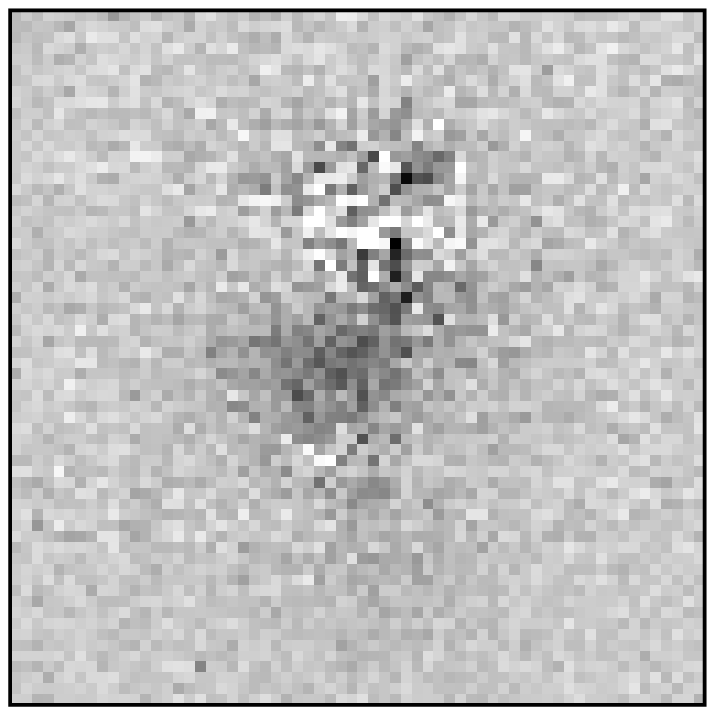}}
\put(75,03){\Large$r'$}
\put(108,00){\includegraphics*[width=34mm]{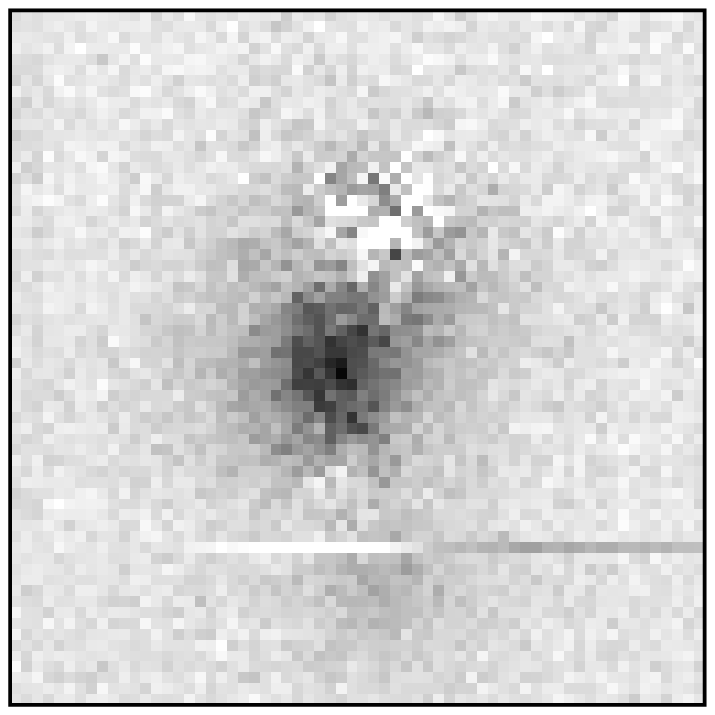}}
\put(111,03){\Large$i'$}
\put(144,00){\includegraphics*[width=34mm]{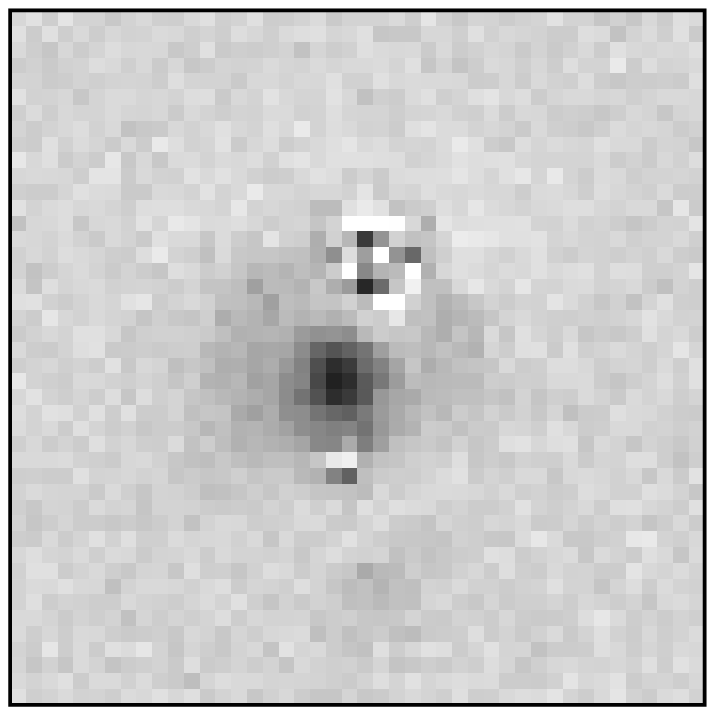}}
\put(147,03){\Large$K_s$}
\end{picture}
\caption[]{Top row: Magellan optical and near-infrared images of the double QSO, 
  in standard astronomical orientation (North up, East left). 
  The size of each image is 4\farcs4.
  The photometric bands are denoted in the lower left corners.
  Bottom row: Corresponding point source-subtracted images, based
  on the best-fit solutions in each band.
  \label{fig:ima}}
\end{figure*}

\begin{table}
\caption[]{Differential astrometry for HE~0047$-$1756, based on the GALFIT results for
the $i$ and $K_s$ band images.
Component A is the brighter of the two QSO components. The quoted error bars 
are only the internal errors given by the fitting routine.
The last object is the faint blob southwest of component B.
\label{tab:pos}}
\begin{tabular}{l.@{\mbox{\hspace{0.5em}$\pm$}\hspace{-1.25em}}.@{\hspace{-1em}}.@{\mbox{\hspace{-2.0em}$\pm$}\hspace{-1.25em}}.}
\hline\noalign{\smallskip}
Component / band & \multicolumn{2}{c}{$\Delta\alpha$} & \multicolumn{2}{c}{$\Delta\delta$} \\
          & \multicolumn{2}{c}{[arcsec]} & \multicolumn{2}{c}{[arcsec]} \\
\noalign{\smallskip}\hline\noalign{\smallskip}
A         &  0.000   & 0.000  &    0.000 \hspace*{3em}   & 0.000 \\[0.5ex]
B ($i$)   &  0.226   & 0.001  &    -1.418 \hspace*{-2em} & 0.001 \\  
B ($K_s$) &  0.230   & 0.001  &    -1.433 \hspace*{-2em} & 0.001 \\[0.5ex]
G ($i$)   &  0.227   & 0.007  &    -0.865 \hspace*{-2em} & 0.013 \\
G ($K_s$) &  0.267   & 0.008  &    -0.881 \hspace*{-2em} & 0.009 \\[0.5ex]
blob      &-0.07 \hspace*{-1em}& 0.15   & -2.36  \hspace*{-1em} & 0.15 \\
\noalign{\smallskip}\hline
\end{tabular}
\end{table}

\begin{table}
\caption[]{Differential photometry for the fainter QSO component and the
lensing galaxy. The numbers are in magnitudes relative to component A.
The last object is again the faint blob southwest of B.
\label{tab:rphot}}
\begin{tabular}{lrrrll}
\hline\noalign{\smallskip}
Component & \multicolumn{1}{c}{$u$} &
            \multicolumn{1}{c}{$g$} &
            \multicolumn{1}{c}{$r$} &
            \multicolumn{1}{c}{$i$} &
            \multicolumn{1}{c}{$K_s$} \\
\noalign{\smallskip}\hline\noalign{\smallskip}
A               & 0.00 & 0.00 & 0.00 & 0.00 & 0.00 \\
B               & 1.31 & 1.39 & 1.49 & 1.50 & 1.46 \\
G               &      &      & 2.23 & 2.10 & 1.44 \\
blob            &      &      &      & 4.8  &      \\
\noalign{\smallskip}\hline
\end{tabular}
\end{table}

\subsection{Optical imaging}

We obtained high resolution optical images at the Baade 6.5~m (Magellan~I)
telescope on 14 Dec 2001, equipped with the \emph{Magellan Instant Camera} (MagIC) 
and a 2k$\times$2k CCD.  The image scale at the f/11 Nasmyth focus was approximately
$0\farcs0692$ per pixel.  The active optics system \citep{schechter*:03:AO} 
updated the focus and translation of the secondary and the twelve most 
flexible elastic modes of the primary mirror at half minute intervals.
A first $r'$ band `snapshot' of 30~sec immediately showed clearly that the
object was double, consisting of at least two point sources.
Three further images were obtained, one each in the SDSS $u'$, $g'$ 
and $i'$ bands
(for simplicity, we shall refer to these bands as \emph{ugri} in the
following). The former was exposed for 300~s, the latter two for 120~s
each. The effective seeing was around $0\farcs 6$, implying that at the image 
scale of $0\farcs 07$/pixel, the images were highly oversampled. The night
was not photometric, however, and we therefore did not attempt to calibrate 
the data with standard stars. Following the usual data reduction steps of
debiasing and flatfielding, we extracted postage stamp images around the QSO 
which are shown in Fig.~\ref{fig:ima}. The clear detection of both components
in the $u$ band is by itself strongly suggestive that both two point sources
show a quasar. We denote the brighter, northern component as A,
and the fainter one as B. 

Relative positions and magnitudes of the individual components
were measured using the program GALFIT kindly provided by
Dr.\ C.Y.~Peng \citep[described in][]{peng*:02:SDG} which performs
simultaneous fits of an arbitrary number of point sources and
extended components to an image. The point spread function (PSF) 
was obtained from the only other reasonably bright star within the 
field of view, about $45''$ NW from the QSO. 
Fitting and subtracting only two scaled PSFs 
gave residuals consistent with zero only for the short wavelength
images. By contrast, the PSF-subtracted $r$ and $i$ images show 
clearly what appears to be the lensing galaxy. 
Including an extended component at the approximate position of the
galaxy, modelled by an de Vaucouleurs law,
considerably improved the fit (total $\chi^2 = 1.34$ per degree
of freedom). The shape of the lensing galaxy, however, is not 
well constrained in the $i$ band image because of blending with
the B component. The astrometric and photometric results of the fits 
are given in Tables \ref{tab:pos}
and \ref{tab:rphot}. The separation between A and B is 
$1\farcs 44$, at a position angle of 261$^\circ$ measured from A.
The coordinates of component A, measured relative to three 
nearby stars listed in the USNOB catalogue, is
R.A. = 00$^\mathrm{h}$~50$^\mathrm{m}$~27\fs83, 
Dec = $-17\degr$~$40'$~$8\farcs8$ ($\pm 0\farcs 1$).

The presence of a galaxy within a few tenths
of an arcsecond of where one would have expected it is compelling
evidence for the lens interpretation.  The faintness of the lensing
galaxy in the filters blueward of $i$ argues for a relatively high
redshift ($z\ga 0.6$) for the lensing galaxy. 

\begin{figure}[tb]
\setlength{\unitlength}{1mm}
\begin{picture}(88,40)
\put(00,00){\includegraphics*[width=40mm]{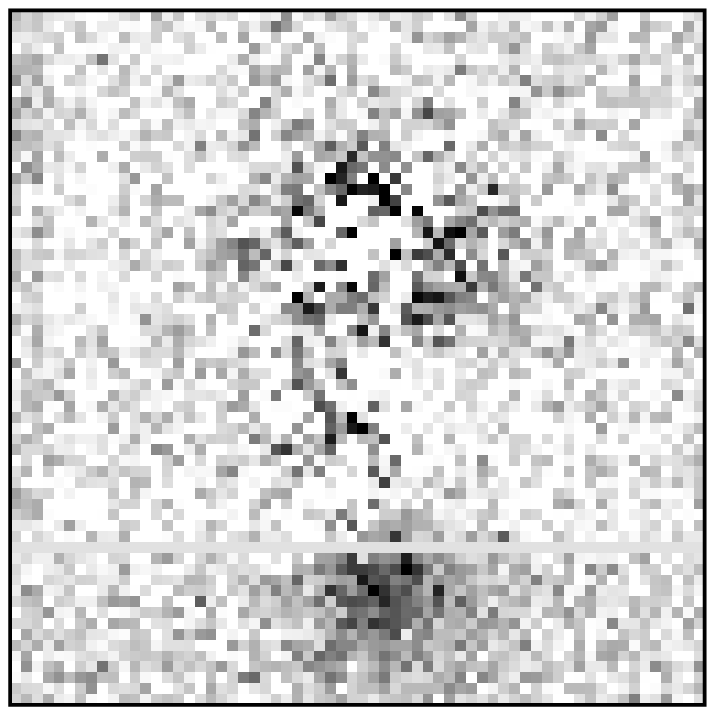}}
\put(00,40){\includegraphics*[width=40mm,angle=-90]{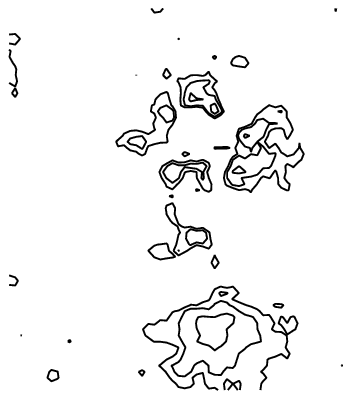}}
\put(03,03){\Large$i'$}
\put(45,00){\includegraphics*[width=40mm]{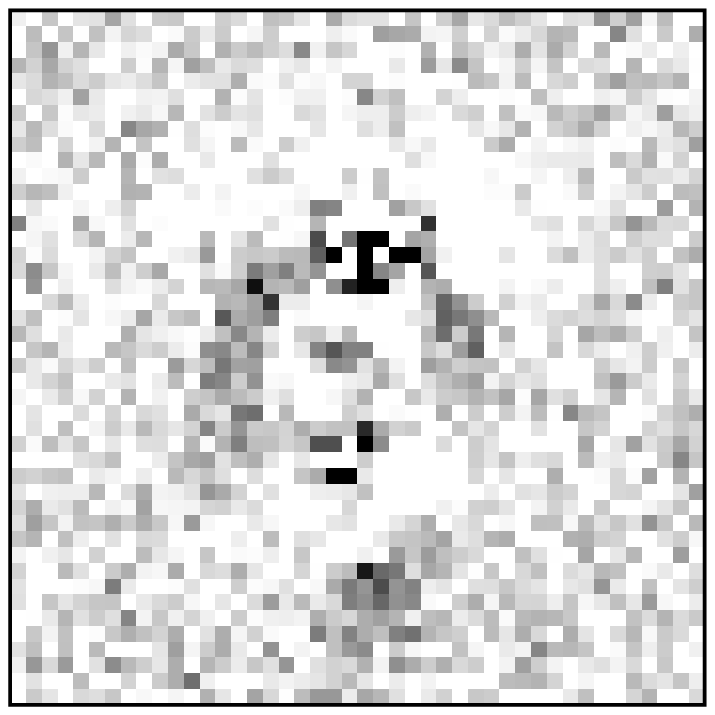}}
\put(45,40){\includegraphics*[width=40mm,angle=-90]{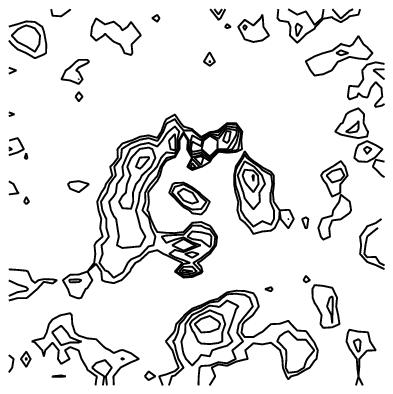}}
\put(48,03){\Large$K_s$}
\end{picture}
\caption[]{Residuals after subtraction of point sources 
  and the lensing galaxy, in the $i$ (left) and $K_s$ (right) 
  bands. Notice the almost complete Einstein ring in the 
  near-infrared image, not present in $i$. 
  \label{fig:ima2}}
\end{figure}

The $i$ band image shows 
also a second, very faint object southwest of component B.
This is more obviously seen in the 
residual image (after PSF and galaxy subtraction) of Fig.~\ref{fig:ima2}.
Approximate position and magnitude estimates ($\pm 0.3$ mag) 
for this faint blob 
have been added to Tables \ref{tab:pos} and \ref{tab:rphot}.

\subsection{Near IR Imaging}

A series of nine dithered 1.5 minute exposures of HE~0047-1756 was
obtained on 2002 November 13 in the $K_s$ filter using CassiCam on the
Baade 6.5-m telescope in non-photometric conditions.  The FWHM in the
stacked image was $0\farcs32$, the pixels are 0\farcs115 on a side.
The lensing galaxy is clearly visible in Fig.~\ref{fig:ima}.  
Simultaneous PSF fitting to the quasar and the galaxy using
GALFIT yielded another set of relative positions and fluxes, 
also listed in Tables \ref{tab:pos} and \ref{tab:rphot}.
Because of the narrower PSF, the $K_s$ band constrains
the shape of the lensing galaxy much better than the optical data.
As seen in the PSF-subtracted image in Fig.~\ref{fig:ima}, the
lensing galaxy is very nearly round. The fitted ellipticity is 
$a/b = 0.81 \pm 0.05$, the half-light radius is 
$0\farcs52 \pm 0\farcs04$.
The differences between the near IR and optical are mostly 
due to uncertainty in the instrument scale.
The position of the lensing galaxy is close to (but not exactly on)
the line connecting components A and B, at a distance
of $0\farcs 89$ from A and $0\farcs 55$ from B.

After subtraction of both PSFs and a model for the lensing galaxy,
the residual image shows significant nonzero flux, most
of which is in form of arc-like features protruding from the 
brighter image and stretching toward the fainter one
(cf.\ Fig.~\ref{fig:ima2}). There
may also be similar, but fainter arcs emanating from component B.
These features are strongly suggestive of an at least partial 
Einstein ring due to the deformed QSO host galaxy. The $i$ band 
image does not show any arcs, consistent with the fact that the 
source is at the redshift of the QSO, in which case the $i$ band
is located below the 4000\,\AA\ break in the source restframe.
Figure \ref{fig:ima2} also shows the companion object
southwest of component B, but only very faintly.

\subsection{Long slit spectroscopy}

Long slit spectra were obtained with the Boller and Chivens
spectrograph on the Baade 6.5-m telescope on 2001 December 21. The
$1''$ slit was oriented North--South, passing through both quasar images.  
Two spectra, of 300~s and 600~s, were taken.  
The two components were spatially well separated.
Both quasar images show emission lines at 4130~\AA\ and 5099~\AA\ 
that we identified as \ion{C}{iv} $\lambda 1550$ and 
\ion{C}{iii}] $\lambda 1909$ at a redshift of 1.67, 
confirming the objective prism redshift. 
The redshift difference $z_B - z_A$ corresponds to a velocity 
difference of less that 30~km~s$^{-1}$, in excellent agreement with
the lensing hypothesis.

\begin{figure}[tb]
\includegraphics*[height=87mm,angle=270]{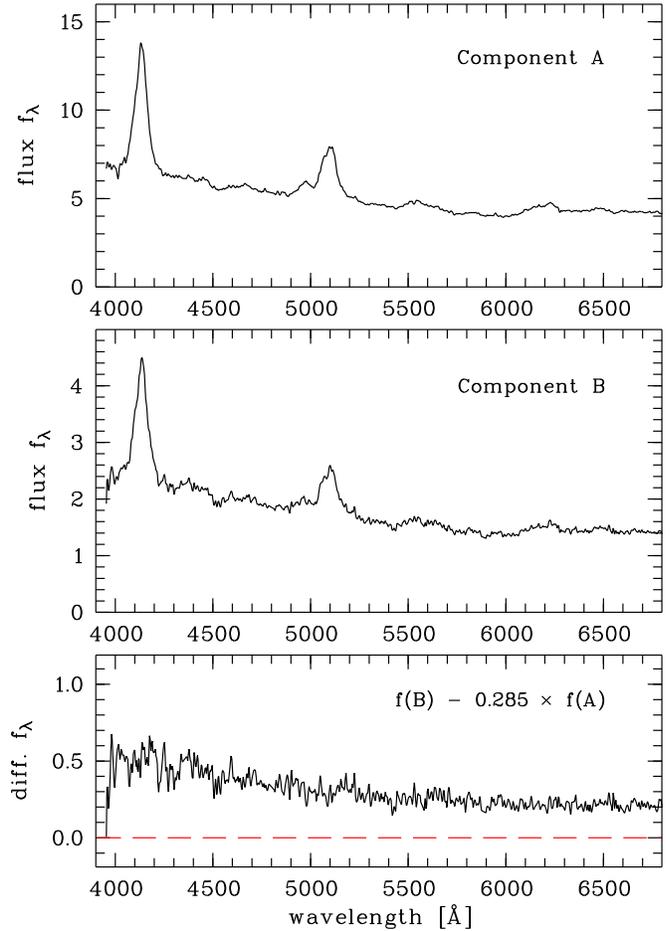}
\caption[]{Top and middle panel: Extracted and calibrated PMAS spectra. Bottom
  panel: Scaled difference spectrum showing the putative microlensing
  component.
   \label{fig:pmasspec}}
\end{figure}

\subsection{Integral field spectrophotometry}

We observed the double QSO on 2002 September 04 with the
Potsdam Multi-Aperture Spectrophotometer PMAS \citep{kelz*:03:PMAS}, 
mounted on the Calar Alto 3.5-m telescope.  The array of $16\times 16$ spatial 
picture elements of each $0\farcs5 \times 0\farcs 5$ was adequate both
to cover the object and to resolve the images. The effective seeing 
was $1\farcs2$, and the night was photometric.
Notice that since the PMAS microlens array reimages 
the exit pupil of the telescope, there are no geometrical losses 
due to incomplete filling of the focal plane.
The data were reduced with the IDL-based software package P3d
\citep{becker:02:D}. The reduction consists of 
standard steps such as debiasing and flatfielding using 
twilight exposures, and dedicated routines such as tracing
and extracting the spectra of individual fibres and reassembling
the data in form of a three-dimensional data cube. The spectra were
extracted with the iterative fitting method described
by \citet{wisotzki*:03:IFS}, which also accounts for differential 
atmospheric refraction. 

In the spectra, shown in Fig.~\ref{fig:pmasspec}, 
we again see \ion{C}{iv} and \ion{C}{iii}] 
as well as \ion{Fe}{ii} features at 5600~\AA\ and 6300~\AA . 
There was insufficient signal to extract a spectrum of the lensing galaxy.
The relative strengths of the emission lines are very similar in the two
components, scaling by a factor of 0.285 (1.36 mag) from A to B.
Subtracting the brighter component, scaled by this factor, from
the fainter one eliminates any trace of the emission lines,
thus also the line profiles are indistinguishable. However, the
continuum does not subtract out by the same factor, and there is
significant residual flux left in component B. Could this excess
be due to contamination by the lensing galaxy? From the spectral
shape of the excess, this would be highly unlikely: We know from
Table~\ref{tab:rphot} that the lensing galaxy is much redder than
the QSO, whereas the excess continuum is markedly blue. Equally,
a contamination from the faint blob south of B can be excluded
given its angular separation; it furthermore does not show up in 
the $g$ band image.

The presence of blue residual flux argues either for differential
microlensing between the two images \citep[very similar to what 
is observed in HE~1104$-$1805,][]{wisotzki*:93:HE1104}, 
or differential temporal variation in the underlying continuum, which is
emitted at two different epochs due to time delay effects. We have
examined the Magellan long-slit spectrum taken 256 days earlier and find
a similar difference with the same sign, although the narrow slit
precludes a quantitative relative spectrophotometry. Nevertheless,
the fact that the flux difference has the same sign at
both spectroscopic epochs favours the microlensing interpretation.  

\section{Models}

A model which works quite well for many lenses is a singular
isothermal sphere with an external shear, as might be generated by the
tide from a neighboring galaxy or cluster.  The two dimensional
projected potential \citep[e.g.,][]{kochanek:91:ILGS} is given by
\[
  \phi(r) = br + {\gamma \over 2} r^2 \cos(\phi - \phi_\gamma)
\]
where $b$ is the diameter of the isothermal sphere in arcseconds, $r$
and $\phi$ are the radial and angular parts, respectively, of the
angular position on the sky with respect to the galaxy, and
$\phi_\gamma$ is the position angle of the shear, measured E of N.
Given our sign convention, a bar or tidal perturber would be at right
angles to $\phi_\gamma$.
Fitting this model to the positions in Table \ref{tab:pos}  and 
to the emission line flux ratio, we get $b=0\farcs770$, $\gamma = 0.0637$ 
and $\phi_\gamma = 104\fdg 9$, with a source position ($\Delta\alpha,
\Delta\delta$) = ($-0\farcs 001, 0\farcs 184$) relative to the measured
position for the lens.  As the number of free parameters is equal to
the number of constraints, the fit is perfect.

The position angle of the shear would imply an object producing a
tide at a position angle of $14\fdg 9$ or $194\fdg 9$.  The faint
blob to the south of the fainter image lies at PA $192\fdg 8$.
The amplitude of the shear is not large.  Taking the blob to
be a singular isothermal sphere at a distance $1\farcs52$ from
the lensing galaxy, the inferred Einstein radius for the blob would be
$0\farcs193$.  This is a factor of 4 smaller than for the lens, implying
a luminosity a factor of 16 smaller if the Faber-Jackson or Tully-Fisher
relation were appropriate. It is consistent with our measurement
(Table~\ref{tab:rphot}) of $i_{\mathrm{blob}} - i_{\mathrm{lens}} = 2.7\pm 0.3$.

The magnifications for the two images 2.1 for the fainter (negative
parity) image and 7.3 for the brighter (positive parity) image. 
The differential time delay is a strong function of the
unknown redshift of the lensing galaxy; it is 32.1 days for
$z_{\mathrm{lens}} = 0.6$ and increases with $z_{\mathrm{lens}}$.

\section{Conclusions}

The newly discovered double QSO HE~0047$-$1756 displays several properties
that make it a showpiece gravitationally lensed object rather than
a binary QSO: The identical (within $\pm 30$~km~s$^{-1}$) 
redshifts of the two quasar images; the similar emission line profiles;
the detection of a galaxy located between the two QSO components;
and the tentative discovery of a partial Einstein ring.

The only apparent deviation from the simple notion of two identical
images is the slight difference in spectral slopes. While spectral
variability and light travel time effects cannot be excluded
as an explanation, we argue that the spectral differences are
more likely due to differential microlensing on the compact
continuum source of the quasar. Spectrophotometric monitoring
is required to resolve the issue unambiguously.

The existence of a prominent NIR Einstein ring makes this object 
particularly interesting for several applications. Firstly,
the geometry of the distortions can be used to obtain additional
constraints on the lens potential, overcoming the shortage of
constraints of double-image lenses. Secondly, the fact that the distorted 
quasar host galaxy is also highly magnified can be exploited to
study the properties of the quasar host. HST imaging data will be
required to accomplish these tasks.

\begin{acknowledgements}
The Hamburg/ESO Survey was supported as ESO key programme 
02-009-45K (145.B-0009). 
MagIC was built with help from a gift by 
Raymond and Beverly Sackler to Harvard University 
and a US NSF grant, AST99-77535, to MIT.
PMAS was partly financed by BMBF/Verbundforschung unter 
053PA414/1 and 05AL9BA1/9.
PLS gratefully acknowleges 
the support of the US NSF under award AST-0206010.
SFS ackowledges financial support provided through
the European Community's Human Potential Program under contract
HPRN-CT-2002-00305, Euro3D RTN.
KJ and LW acknowledge a DFG travel grant under Wi 1369/12-1.

\end{acknowledgements}

\bibliographystyle{aa}
\bibliography{lutzbib,aipbib,ownpubrefjourn,ownpubproceed}

\begin{thebibliography}{10}
\expandafter\ifx\csname natexlab\endcsname\relax\def\natexlab#1{#1}\fi

\bibitem[{Becker(2002)}]{becker:02:D}
Becker, T. 2002, PhD thesis, Universit{\"a}t Potsdam

\bibitem[{{Kelz} {et~al.}(2003){Kelz}, {Roth}, \& {Becker}}]{kelz*:03:PMAS}
{Kelz}, A., {Roth}, M.~M., \& {Becker}, T. 2003, Proc.~SPIE, 4841, 1057

\bibitem[{{Kochanek}(1991)}]{kochanek:91:ILGS}
{Kochanek}, C.~S. 1991, \apj, 373, 354

\bibitem[{{Peng} {et~al.}(2002){Peng}, {Ho}, {Impey}, \& {Rix}}]{peng*:02:SDG}
{Peng}, C.~Y., {Ho}, L.~C., {Impey}, C.~D., \& {Rix}, H. 2002, \aj, 124, 266

\bibitem[{{Schechter} {et~al.}(2003){Schechter}, {Burley}, {Hull}, {Johns},
  {Martin}, {Schaller}, {Shectman}, \& {West}}]{schechter*:03:AO}
{Schechter}, P.~L., {Burley}, G.~S., {Hull}, C.~L., {et~al.} 2003, Proc.~SPIE,
  4837, 619

\bibitem[{Turner(1980)}]{turner:80:GLQ}
Turner, E.~L. 1980, ApJ, 242, L135

\bibitem[{{Wisotzki} {et~al.}(2003){Wisotzki}, {Becker}, {Christensen},
  {Helms}, {Jahnke}, {Kelz}, {Roth}, \& {Sanchez}}]{wisotzki*:03:IFS}
{Wisotzki}, L., {Becker}, T., {Christensen}, L., {et~al.} 2003, A\&A, 408, 455

\bibitem[{Wisotzki {et~al.}(2000)Wisotzki, Christlieb, Bade, Beckmann,
  {K\"ohler}, Vanelle, \& Reimers}]{wisotzki*:00:HES3}
Wisotzki, L., Christlieb, N., Bade, N., {et~al.} 2000, A\&A, 358, 77

\bibitem[{Wisotzki {et~al.}(1993)Wisotzki, K\"ohler, Kayser, \&
  Reimers}]{wisotzki*:93:HE1104}
Wisotzki, L., K\"ohler, T., Kayser, R., \& Reimers, D. 1993, A\&A, 278, L15

\bibitem[{{Wisotzki} {et~al.}(2002){Wisotzki}, {Schechter}, {Bradt}, {Heinm{\"
  u}ller}, \& {Reimers}}]{wisotzki*:02:HE0435}
{Wisotzki}, L., {Schechter}, P.~L., {Bradt}, H.~V., {Heinm{\" u}ller}, J., \&
  {Reimers}, D. 2002, A\&A, 395, 17

\end{thebibliography}

\end{document}